\documentclass[twocolumn]{aastex63_mod}

\usepackage{xspace}
\usepackage{amsmath}
\usepackage{hyperref}
\newcommand{\msun}{\hbox{$M_\odot$}\xspace}
\newcommand{\msyr}{\hbox{$M_\odot~\text{yr}^{-1}$}\xspace}

\newcommand{\kmsmpc}{\hbox{$\text{km s}^{-1} \text{ Mpc}^{-1}$}\xspace}

\newcommand{\prosp}{\texttt{Prospector}\xspace}

\usepackage{booktabs}
\usepackage{changepage}

\date{\today}
\shorttitle{Empirical Dust Attenuation Model Improves Simulated UVJ Diagram}
\shortauthors{Nagaraj et al.}
\graphicspath{{./}{figures/}}

\begin{document}

\title{Empirical Dust Attenuation Model Leads to More Realistic UVJ Diagram for TNG100 Galaxies}

\correspondingauthor{Gautam Nagaraj}
\email{gxn75@psu.edu}

\author[0000-0002-0905-342X]{Gautam Nagaraj}
\affil{Department of Astronomy \& Astrophysics, The Pennsylvania State University, University Park, PA 16802, USA}
\affil{Institute for Gravitation and the Cosmos, The Pennsylvania State University, University Park, PA 16802, USA}
\affiliation{Center for Computational Astrophysics, 162 Fifth Avenue, New York, NY, 10010, USA}

\author[0000-0002-1975-4449]{John C. Forbes}
\affiliation{Center for Computational Astrophysics, 162 Fifth Avenue, New York, NY, 10010, USA}

\author[0000-0001-6755-1315]{Joel Leja}
\affil{Department of Astronomy \& Astrophysics, The Pennsylvania State University, University Park, PA 16802, USA}
\affil{Institute for Gravitation and the Cosmos, The Pennsylvania State University, University Park, PA 16802, USA}
\affil{Institute for Computational \& Data Sciences, The Pennsylvania State University, University Park, PA 16802, USA}

\author[0000-0002-9328-5652]{Dan Foreman-Mackey}
\affiliation{Center for Computational Astrophysics, 162 Fifth Avenue, New York, NY, 10010, USA}

\author[0000-0003-4073-3236]{Christopher C. Hayward}
\affiliation{Center for Computational Astrophysics, 162 Fifth Avenue, New York, NY, 10010, USA}

\begin{abstract}

Dust attenuation varies substantially from galaxy to galaxy and as of yet cannot be reproduced from first principles in theoretical models. In \cite{Nagaraj2022a}, we developed the first Bayesian population model of dust attenuation as a function of stellar population properties and projected galaxy shape, built on spectral energy distribution (SED) fits of nearly 30,000 galaxies in the 3D-HST grism survey with broadband photometric coverage from the rest-frame UV to IR. In this paper, we apply the model to galaxies from the large-volume cosmological simulation TNG100. We produce a UVJ diagram and compare it with one obtained in previous work by applying approximate radiative transfer to the simulated galaxies. We find that the UVJ diagram based on our empirical model is in better agreement with observations than the previous effort, especially in the number density of dusty star forming galaxies. We also construct the intrinsic dust-free UVJ diagram for TNG and 3D-HST galaxies at $z\sim 1$, finding qualitative agreement but residual differences at the $10-20\%$ level. These differences can be largely attributed to the finding that TNG galaxies have, on average, $29\%$ younger stellar populations and $0.28$ dex higher metallicities than observed galaxies.



\end{abstract}


\section{Introduction} \label{sec:intro}

Dust plays a major role in shaping the ultraviolet (UV) through infrared (IR) portion of galaxy spectra, absorbing and scattering UV through near-IR (NIR) light and re-radiating it in the mid- and far-IR \citep[MIR and FIR, e.g.,][]{Draine2003rev}. As dust preferentially absorbs and scatters bluer light, it causes not only extinction but also reddening.

Due to the small angular size covered by cosmologically distant galaxies, we typically are able to observe only integrated photometry and spectroscopy, rather than resolved lines of sight. In a galaxy, the complex configuration of stars and dust causes differential levels of obscuration, compounded by scattering both into and out of different lines of sight. By re-imagining the scenario as a dust screen (i.e., a sheet) in front of the stars \citep[e.g.,][]{Calzetti1994,Calzetti2000,Chevallard2013}, we can create an effective dust extinction law for the galaxy, designated as dust attenuation.

Dust attenuation has been shown to vary widely from galaxy to galaxy \citep[e.g.,][]{Burgarella2005,Noll2009,Wild2011,Buat2012,Arnouts2013,Kriek2013,Reddy2015,Salim2016,Leja2017,Salim2018}. For a given galaxy, dust attenuation is difficult to measure given the similar reddening effects it has on the galaxy spectrum as increasing metallicity or average stellar age \citep[e.g.,][]{Conroy2013,Santini2015}. In fact, dust attenuation is a major source of uncertainty in high redshift star formation studies \citep[e.g.,][]{Bouwens2012,Finkelstein2012,Oesch2013}.

While our understanding of dust grains in the local universe is substantial \citep[][and references therein]{GGJ2018}, the reduced level of knowledge at high redshift coupled with the endless intricate possibilities of star-dust geometry severely limits our theoretical foundations for dust attenuation in the early universe. Therefore, theoretical efforts in understanding galaxy evolution, from analytic models \citep[e.g.,][]{Forbes2014,Forbes2019} to hydrodynamical simulations \citep[e.g.,][]{Genel2014,Pillepich2018,Dave2019}, often exclude dust physics. For such models, an empirically based dust attenuation prescription written as a function of physical properties of galaxies such as stellar mass, star formation rate (SFR), metallicity, and redshift would be extremely useful.

Many astronomers have explored the connection between dust attenuation curves and galaxy properties, both in the local universe \citep[e.g.,][]{Burgarella2005,Salim2016,Leja2017,Salim2018} and at high redshift \citep[e.g.,][]{Arnouts2013,Kriek2013,Salmon2016,Tress2018}. Nevertheless, a complete understanding of dust attenuation is still far from realization \citep[][and references therein]{SalimNarayanan2020}.

\cite{Nagaraj2022a}, hereafter referred to as Paper~I, describes the first Bayesian population model for dust attenuation (as a function of stellar population properties and projected galaxy shape) developed from state-of-the-art spectral energy distribution (SED) fits \citep[Prospector,][]{Leja2017,Leja2019,Leja2020} of a mass-complete set of $\sim 29,000$ $0.5<z<3.0$ galaxies from the 3D-HST grism survey \citep{Brammer2012,Skelton2014,Momcheva2016}. By combining the grism redshifts with copious multiwavelength photometry from CANDELS \citep{Grogin2011,Koekemoer2011}, the 3D-HST survey allows for more accurate inferred galaxy properties. The multi-layered aspect of the model properly accounts \citep{ForemanMackey2014HierBay} for the large uncertainties and covariances among galaxy parameters \citep[e.g.,][]{Conroy2013,Santini2015}, making the model more robust against complex correlations between inferred galaxy properties. We demonstrate this improved accuracy in a test in Paper~I.

In this paper, we highlight how to use the (unresolved) population model and demonstrate that it is more effective at reproducing the observed galaxy colors than a first-principles approach \citep{Nelson2018,Donnari2019} by applying it to subhalos (galaxies) from the IllustrisTNG simulations \citep{Pillepich2018} at redshift $z=1$ and comparing the $U-V$ and $V-J$ colors we derive to those of \cite{Donnari2019} and to an observed set of colors from 3D-HST. Having an accurate working model for dust attenuation allows us to identify specific areas of improvement for theoretical models of galaxy evolution by comparing to observations since cosmological simulations are not yet at a level to reproduce dust attenuation from first principles.

In \S \ref{sec:uhm}, we present a visualization of the model to show how it can be directly applied to theoretical models to produce attenuated spectra. In \S \ref{sec:methods}, we introduce the UVJ diagram and discuss how we calculated colors for TNG100 galaxies. Then, in \S \ref{sec:res} we present the UVJ colors of TNG100 and compare them to both previous results from \cite{Donnari2019} and the measured 3D-HST colors. We discuss our results in \S \ref{sec:disc} and conclude in \S \ref{sec:conc}.

To ensure consistency with \cite{Nelson2018} and \cite{Donnari2019}, we use the cosmological parameters from \cite{PlanckCosParam2016}, namely $\Omega_{m,0}=0.3089$, $\Omega_{\Lambda,0}=0.6911$, $\Omega_{b,0}=0.0486$, and $H_0=67.74$ \kmsmpc; the \cite{Chabrier2003} IMF; and the AB magnitude system \citep{Oke1974}.  

\section{Using the Population Model} \label{sec:uhm}

Our population models of dust attenuation\footnote{https://github.com/Astropianist/DustE}, as described in detail in Paper~I, describe both the two-component dust attenuation curve of \prosp \citep{Leja2017,Johnson2021}, inspired by \cite{CharlotFall2000}, and an effective attenuation curve (single component). In the two-component model, the birth cloud dust attenuation ($\tau_{\lambda,1}$), which affects only stars less than 10 million years old, is modeled as $\tau_{\lambda,1} \propto \lambda^{-1}$. The diffuse dust attenuation ($\tau_{\lambda,2}$) is formulated according to \cite{Noll2009} and \cite{Kriek2013}. This is a flexible extension of the attenuation curve from \cite{Calzetti2000}, or ``Calzetti Law'' (represented as $k'(\lambda)$ below) with parameters for slope and 2175 \AA~bump strength. \prosp fixes the bump strength $E_b$ based on results from \cite{Kriek2013}. 

\begin{equation}
    \tau_{\lambda,1} = \tau_1 \left( \frac{\lambda}{5500 \AA} \right)^{-1}
\end{equation}

\begin{equation} \label{eq:diffdust}
    \tau_{\lambda,2} = \frac{\tau_2}{4.05}\left(k'(\lambda)+D(\lambda) \right) \left( \frac{\lambda}{5500 \AA} \right)^{n}
\end{equation}

\begin{equation}
    D(\lambda) = \frac{E_b(\lambda \Delta \lambda)^2}{(\lambda^2-\lambda_0^2)^2 + (\lambda \Delta \lambda)^2}
\end{equation}

\begin{equation}
    \lambda_0,~\Delta \lambda,~E_b = 2175 \AA,~350 \AA,~0.85-1.9n
\end{equation}

The parameters fitted in the two-component model are $n$ (related to slope), $\tau_2$, and $\tau_1$. In the effective dust attenuation model, the curve is parameterized by Equation \ref{eq:diffdust}, with parameters $n_{\rm eff}$ and $\tau_{\rm eff}$. In this work, we use the two-component dust attenuation model, in which $\tau_1$ is a linear interpolation function of $\tau_2$ while $n$ and $\tau_2$ are fitted as a linear interpolation function of stellar mass, SFR, stellar metallicity, redshift, and axis ratio (a proxy for inclination). Both are models described in Paper~I.



The package we have created to access the population dust models is quite simple to use, and we provide a number of demos in the footnote link given earlier. A description of the code is also presented in Paper~I.

\begin{figure*}
    \centering
    \resizebox{\hsize}{!}{
    \includegraphics{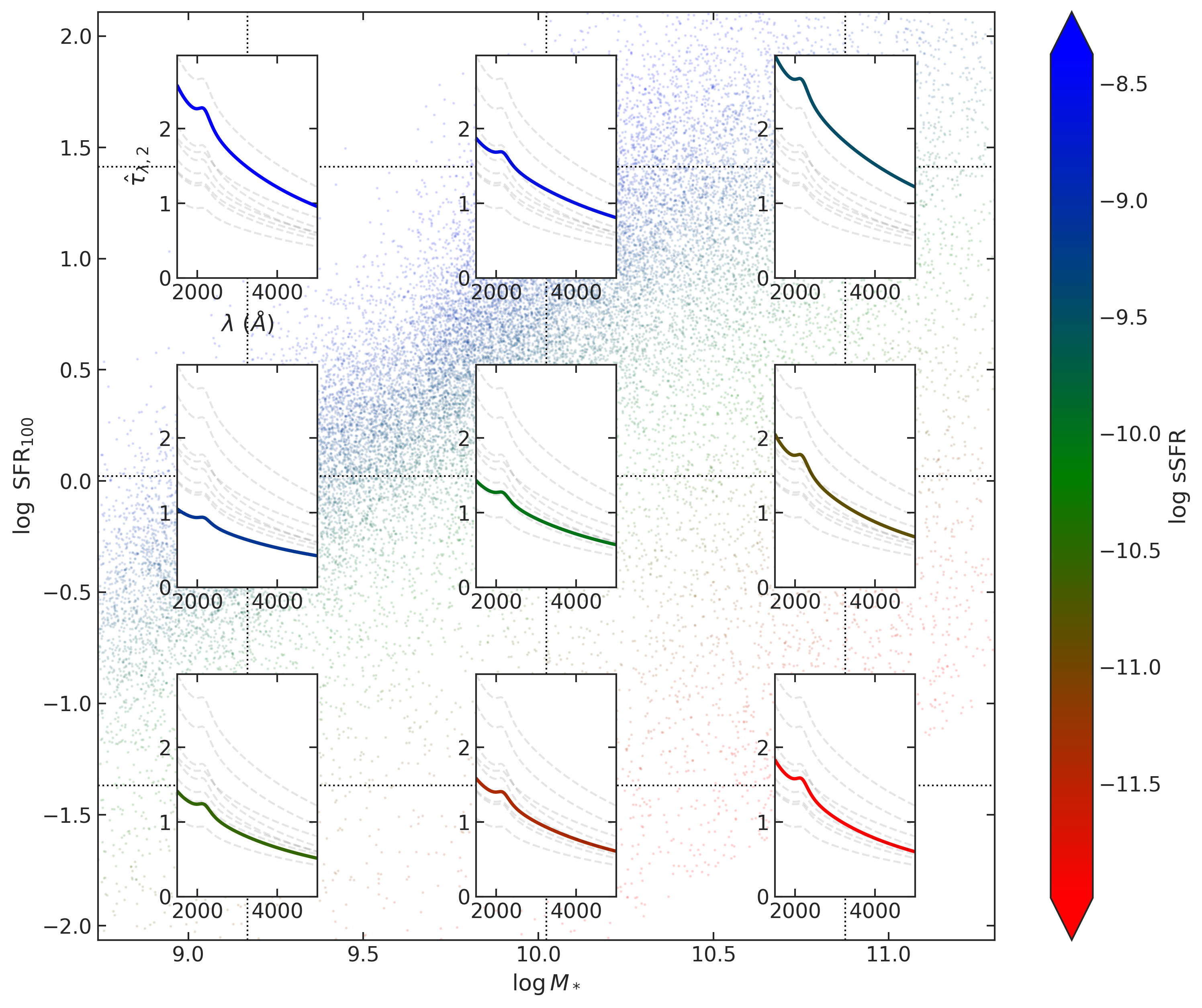}}
    \caption{Variation in the diffuse dust attenuation curve over the stellar mass -- SFR parameter space. The attenuation curves correspond to the model's prediction at the location in parameter space at the center of each small plot (as highlighted by the dotted black lines). The $1-\sigma$ errors on the mean attenuation curves are shaded but are too small to be visible. The faint points in the background show the mean \prosp posterior sample for stellar mass and SFR and are colored by sSFR, as are the attenuation curves themselves. For easier interpretation, in each attenuation curve plot, we show the curve for the other eight locations as gray dashed lines. We can see that attenuation generally increases (larger optical depth and typically steeper attenuation slope) with stellar mass and SFR, but the detailed behavior is rather complex. See Paper~I for extensive discussions of the science results of our models.}
    \label{fig:attncurves}
\end{figure*}

As a demonstration of our models, Figure \ref{fig:attncurves} shows the evolution of the diffuse dust attenuation curve over the stellar mass -- SFR parameter space. We observe that the attenuation generally increases (larger optical depth and typically steeper slope) with stellar mass and SFR, but the exact relation between the curves and the parameter space are quite complex. Discussions on the scientific implications of our models can be found in Paper~I. 

Also, in Paper~I, using mock tests, we show that the population model is substantially more accurate than the common procedure of binning in, or even fitting a Bayesian model to, point estimates of each galaxy's properties. By averaging the likelihood over the full posterior of each object, we can avoid our results being driven by well-known correlations between parameters in individual fits (e.g., $\tau_2$ and $n$) and thus produce more reliable inferred relationships. With 5-D dust attenuation models, there is a plethora of interesting trends that are still to be gleaned, making the package we provide useful for learning more about dust attenuation in addition to the purpose it serves in providing an interface between theory and observations as demonstrated in this work.

\section{Methods} \label{sec:methods}
\subsection{Introduction to the UVJ Diagram} \label{subsec:uvj_intro}

It is important to be able to easily distinguish between galaxies that are red due to quiescence and those that are red due to dust. One simple observational technique is to place the galaxies on a rest-frame $U-V$ vs.\ $V-J$ color-color plot \citep{Wuyts2007}. While a single color ($U-V$ or $V-J$) produces a degeneracy between intrinsically red objects and systems that have been reddened by dust, \cite{Wuyts2007} showed that the two galaxy classes separate in this two-color space, as dusty star forming galaxies tend to have redder $V-J$ colors than their quiescent counterparts. Many studies have used rest-frame $UVJ$ colors to distinguish quiescent and star-forming galaxies \citep[e.g.,][]{Williams2009,Patel2012,Whitaker2012,Muzzin2013uvj,Fumagalli2014,Bowman2019}. 

More fundamentally, quiescence can be defined via specific SFR (sSFR), though observational inferences of sSFR are more subject to systematic effects than rest-frame colors. Of course, in simulations, the true sSFR is known. However, there is no precise definition for quiescence, and sSFR limits range from $10^{-10}$ yr$^{-1}$ \citep[e.g.,][]{Whitaker2017,Wu2018} to $10^{-11}$ yr$^{-1}$ \citep[e.g.,][]{Cecchi2019}. Nevertheless, \cite{Leja2019b} have shown that the standard UVJ cuts correspond to $\log~{({\rm sSFR/yr}^{-1})}\sim -10.0$. 


\subsection{Measuring Magnitudes for TNG100 Subhalos} \label{subsec:meastng}

To exhibit our population dust attenuation models' validity, we apply them to simulated galaxies from the IllustrisTNG project\footnote{\url{https://www.tng-project.org/}} \citep{Pillepich2018}, namely TNG100 \citep{Marinacci2018,Naiman2018,Nelson2018,Pillepich2018b,Springel2018}, and compare the derived UVJ colors to those of \cite{Donnari2019} and the observations of 3D-HST galaxies from \prosp \citep{Leja2019,Leja2020}. The results are shown in \S \ref{sec:res}. 


\cite{Donnari2019} apply the \cite{Nelson2018} resolved dust model C to TNG simulated galaxies and measure the magnitudes $U$, $V$, and $J$ of the resulting spectra. To ensure a suitable comparison to their work, we employed a similar process to that outlined in \cite{Nelson2018} and \cite{Donnari2019} to measure colors of TNG galaxies, with the major exception being the treatment of dust. We took all TNG100 $z=1$ (snapshot 50) subhalos with stellar masses of at least $10^9 \msun$ and summed up the magnitudes in the U and V \citep{Bessell1990} and J \citep{Skrutskie2006} filters for the stellar particles within $2r_{1/2}$ in each subhalo, where $r_{1/2}$ is the half-mass radius. 

To model spectra, we used the FSPS stellar population synthesis code \citep{Conroy2009,Conroy2010SPSM,ForemanMackey2014} with a Chabrier IMF \citeyear{Chabrier2003}, the Padova isochrones, and the MILES spectral library. To speed up the computation process, we calculated U, V, and J magnitudes using FSPS for each of the metallicity and age steps in the Padova database. We treated all of these grid points as simple stellar populations (single-age, single-metallicity populations). We included nebular emission \citep{Byler2017}, with the gas-phase metallicity equal to the stellar metallicity for self consistency and with ionization parameter fixed at $\log~U=-2.0$. 

Again following the steps outlined in \cite{Nelson2018} and \cite{Donnari2019}, for each stellar particle within $2r_{1/2}$ (twice the stellar half-mass radius) of a given subhalo, we used a bicubic spline over log metallicity and log age to approximate the magnitudes and then added up the fluxes in proportion to the mass of each particle. Metallicities and ages outside the grid bounds yield the magnitudes at the nearest boundary point.


The aforementioned process measures dust-free magnitudes for the stellar populations in each subhalo. To include dust, we looked up the attenuation parameters of our population model given the values of stellar mass, SFR, metallicity, redshift, and axis ratio $b/a$ (proxy for inclination) for each subhalo. We used the average SFR over the last 100 million years of the star formation history (SFH) and the mass-weighted metallicity for the calculation to emulate observations of each quantity. 

Axis ratio depends on viewing angle, which is arbitrary for the simulated galaxies, so we used the following procedure to select axis ratios for TNG galaxies. The connection between axis ratio and inclination becomes tenuous for galaxies with triaxial morphologies \citep{vanderwel2014b}. Galaxies with lower SFRs and/or stellar masses tend to fall in this category. Therefore, the values chosen for axis ratio in this regime do not matter. Thus, for galaxies with $\log({\rm SFR}/[\msyr])<1$, we simply drew values for $b/a$ from a uniform distribution between $0$ and $1$. 

For galaxies with $\log({\rm SFR}/[\msyr])\geq 1$, we employed the simple triaxial model from \cite{Zuckerman2021}. The sources are treated as ellipsoids with axes parameters $a'$, $b'=a'$, and $c'$ (thickness). We let the ratio of thickness $c'$ to the semi-major axis $a'$ be $D \equiv c'/a' = 0.1$. Then, for a given inclination $\phi$, where $0$ corresponds to face-on, the relationship between axis ratio $b/a$ and inclination $\phi$ is given by the following formula.

\begin{equation}
    \frac{b}{a} = \frac{D}{\sqrt{\sin^2{\phi} + D^2 \cos^2{\phi}}}
\end{equation}

We drew values of $\cos{\phi}$ from a uniform distribution between $[0,1]$ (consistent with a random viewing angle) and calculated axis ratio $b/a$ for each inclination.


We corrected the measured U, V, and J magnitudes of stellar particles for dust attenuation and then added the fluxes of particles within $2r_{1/2}$ to get the overall colors for each subhalo. 


\subsection{Measuring Colors for 3D-HST Galaxies} \label{subsec:meas3dhst}

To measure rest-frame colors for observed galaxies at cosmological distances, we need a predicted spectrum conditioned on existing data. This means that even observationally-inferred rest-frame UVJ-colors are model-dependent. 

In our case, we are using spectra generated through FSPS during \prosp fits of 3D-HST galaxies. \prosp uses nested sampling methods \citep{Speagle2020} to efficiently probe the physical parameter space in order to create a spectrum that best fits the observed photometry. 


This process is complicated by degeneracies between parameters such as age, metallicity, and dust attenuation \citep[e.g.,][]{Conroy2013,Santini2015}. As changes in these parameters have similar effects on the generated spectrum, we can have different parameter combinations with the same rest-frame colors. 


In this work, we also compare dust-free TNG colors to the inferred rest-frame UVJ colors when dust is removed to directly study properties of TNG galaxies without the added layer of our dust model. For the measurement of dust-free colors, the degeneracies between dust, age, and metallicities become particularly important as we set the dust attenuation to zero. 

In order to reduce the effects of degeneracies and generate more reliable dust-free colors, we use the dust model from Paper~I hierarchically to re-weight the \prosp posterior samples of individual galaxies. This process is called hierarchical shrinkage: we use what we have learned about dust attenuation for the entire sample to help restrict the parameter space for individual galaxies. 


We now derive the weights used in the process described above. We consider a single galaxy. The physical parameters that influence dust attenuation (i.e., stellar mass, SFR, stellar metallicity, redshift, and axis ratio, in our model) are denoted as $\mathbf{w}$. Meanwhile, the parameters of our dust model for all galaxies in the sample are called $\boldsymbol{\theta}$. 

Next, we label the (fixed) parameters defining the \prosp priors as $\boldsymbol{\alpha}$. While most priors are flat over the parameters of interest, the exact limits of those priors belong to $\boldsymbol{\alpha}$. In addition, there are informative priors for some variables, including diffuse dust optical depth $\tau_2$. For $\tau_2$, the mean, standard deviation, and limits of the truncated normal distribution serving as the prior are part of $\boldsymbol{\alpha}$. We denote the set of all photometric observations for the galaxy $\mathbf{X}$. In all equations below $p(\ldots)$ represents the probability distribution of the quantity in the parentheses.

We can then write the following formula for the expected value of a function $f$ for the galaxy. In our case, $f$ is the value of the dust attenuation parameters $n$ and $\tau$ as a function of $\mathbf{w}$ given population parameters $\boldsymbol{\theta}$. We show that $f$ depends on the particular population parameters $\boldsymbol{\theta}$ using a subscript: $f_{\boldsymbol{\theta}}$.

\begin{equation} \label{eq:expval}
    \langle f_{\boldsymbol{\theta}}\rangle = \int f_{\boldsymbol{\theta}}(\mathbf{w})p(\mathbf{w}|\mathbf{X},\boldsymbol{\theta}) d\mathbf{w}
\end{equation}

In this equation, we have assumed that the galaxy observations $\mathbf{X}$ do not depend directly on the population model parameters $\boldsymbol{\theta}$. Next, from Bayes' Theorem, we can write the following equation.

\begin{equation} \label{eq:bayes}
    p(\mathbf{w}|\mathbf{X},\boldsymbol{\Omega}) = \frac{p(\mathbf{X}|\mathbf{w}) p(\mathbf{w}|\boldsymbol{\Omega})}{p(\mathbf{X}|\boldsymbol{\Omega})}
\end{equation}

In Equation \ref{eq:bayes}, we have made the reasonable assumption that $\mathbf{X}$ does not depend directly on the interim priors $\boldsymbol{\alpha}$, nor, again, on the population parameters $\boldsymbol{\theta}$, but only on the galaxy properties $\mathbf{w}$. Let $\boldsymbol{\Omega}$ represent either quantity, so this assumption is equivalent to  $p(\mathbf{X}|\boldsymbol{\Omega},\mathbf{w}) = p(\mathbf{X}|\mathbf{w})$. Using Equation \ref{eq:bayes}, we can substitute for $p(\mathbf{w}|\mathbf{X},\boldsymbol{\theta})$ and multiply the integrand in Equation \ref{eq:expval} by 1 in the following manner.

\begin{equation}
    \langle f_{\boldsymbol{\theta}}\rangle = \int f_{\boldsymbol{\theta}}(\mathbf{w}) \frac{p(\mathbf{X}|\mathbf{w}) p(\mathbf{w}|\boldsymbol{\theta})}{p(\mathbf{X}|\boldsymbol{\theta})} \left[ \frac{p(\mathbf{w}|\mathbf{X},\boldsymbol{\alpha}) p(\mathbf{X}|\boldsymbol{\alpha})}{p(\mathbf{w}|\boldsymbol{\alpha}) p(\mathbf{X}|\mathbf{w})} \right] d\mathbf{w}
\end{equation}

\noindent Therefore, we end up with the simplified result below.

\begin{equation} \label{eq:simp1}
    \langle f_{\boldsymbol{\theta}}\rangle = \int f_{\boldsymbol{\theta}}(\mathbf{w}) \frac{p(\mathbf{w}|\boldsymbol{\theta}) p(\mathbf{X}|\boldsymbol{\alpha})}{p(\mathbf{w}|\boldsymbol{\alpha}) p(\mathbf{X}|\boldsymbol{\theta})} p(\mathbf{w}|\mathbf{X},\boldsymbol{\alpha}) d\mathbf{w}
\end{equation}

For a given set of parameters $\boldsymbol{\theta}$, the quantity $\frac{p(\mathbf{X}|\boldsymbol{\alpha})}{p(\mathbf{X}|\boldsymbol{\theta})}$ is constant over $\mathbf{w}$ and therefore can be moved outside the integral. We hereafter refer to this quantity as $1/Z_{\boldsymbol{\theta}}$. For convenience, we label $\frac{p(\mathbf{w}|\boldsymbol{\theta})}{p(\mathbf{w}|\boldsymbol{\alpha})}$ as $g_{\boldsymbol{\theta}}(\mathbf{w})$. We thus rewrite Equation \ref{eq:simp1} as the following.

\begin{equation}
    \langle f_{\boldsymbol{\theta}}\rangle = \frac{1}{Z_{\boldsymbol{\theta}}} \int f_{\boldsymbol{\theta}}(\mathbf{w}) g_{\boldsymbol{\theta}}(\mathbf{w}) p(\mathbf{w}|\mathbf{X},\boldsymbol{\alpha}) d\mathbf{w}
\end{equation}

\noindent If we let the function $f$ be unity, the integral above must yield a value of $1$. Therefore, we find that the normalization constant $Z_{\boldsymbol{\theta}}$ is given by the following formula.

\begin{equation}
    Z_{\boldsymbol{\theta}} = \int g_{\boldsymbol{\theta}}(\mathbf{w}) p(\mathbf{w}|\mathbf{X},\boldsymbol{\alpha}) d\mathbf{w}
\end{equation}

Since we have at our disposal posterior samples from \prosp fits of 3D-HST galaxies, which are sampled from the distribution $\mathbf{w}_n \sim p(\mathbf{w}|\mathbf{X},\boldsymbol{\alpha})$, we can use the Monte Carlo integration technique to approximate the integrals above. If we have $N$ posterior samples for a given galaxy, we get the following result.

\begin{equation} \label{eq:expvalapprox}
    \langle f_{\boldsymbol{\theta}}\rangle \approx \frac{1/N \sum_{n=1}^N f_{\boldsymbol{\theta}}(\mathbf{w}_n) g_{\boldsymbol{\theta}}(\mathbf{w}_n) }{1/N \sum_{n=1}^N g_{\boldsymbol{\theta}}(\mathbf{w}_n)}
\end{equation}

Equation \ref{eq:expvalapprox} gives a straightforward method to measure the weight for any given \prosp posterior sample and set of population model parameters. 

Ultimately, we want to get the expected value of $f$ not just conditioned on any one value of the population parameters $\boldsymbol{\theta}$, but rather the full posterior distribution of $\boldsymbol{\theta}$ conditioned simultaneously on the photometry from all $\sim 30,000$ 3D-HST galaxies which we obtained in Paper~I. In that case, we have another integral over $d\boldsymbol{\theta}$ and can do another Monte Carlo integration using the posterior samples from the population modeling efforts to find the effective weights. 

By applying the population-model-posterior-averaged weights to the \prosp posterior samples, we measure values of $n$ and $\tau$ for each galaxy that are less impacted by parameter correlations in the SED fitting process. Thus, we are able to compute more reliable dust-attenuated and especially dust-free UVJ colors. 

In Section \ref{sec:res}, we compare data to simulations in both the dust-free and with-dust spaces (reverse and forward modeling, respectively). In order to calculate dust-free 3D-HST colors, we use the equations above to reweight the \prosp posterior samples for each galaxy by our hierarchical population model, take the weighted average of the samples, set the dust attenuation parameters to zero, and then run FSPS to get the resulting spectrum. For comparisons with the simulated galaxies, we limit the redshift range to $0.8\leq z \leq 1.2$ to better mirror the $z=1$ selection in TNG. We also impose the mass-completeness limits in the 3D-HST sample on the TNG galaxies (at $z=1$).

\section{TNG100 vs 3D-HST UVJ Diagrams} \label{sec:res}

We first compare dust-free colors of our TNG100 subhalos and \prosp dust-free rest-frame colors conditioned on 3D-HST photometry. There is scatter between the TNG dust-free colors measured by \cite{Donnari2019} and ours on the order of $0.05$ magnitudes with negligible bias. These do not affect our conclusions.

\begin{figure*}
    \centering
    \includegraphics[scale=0.5]{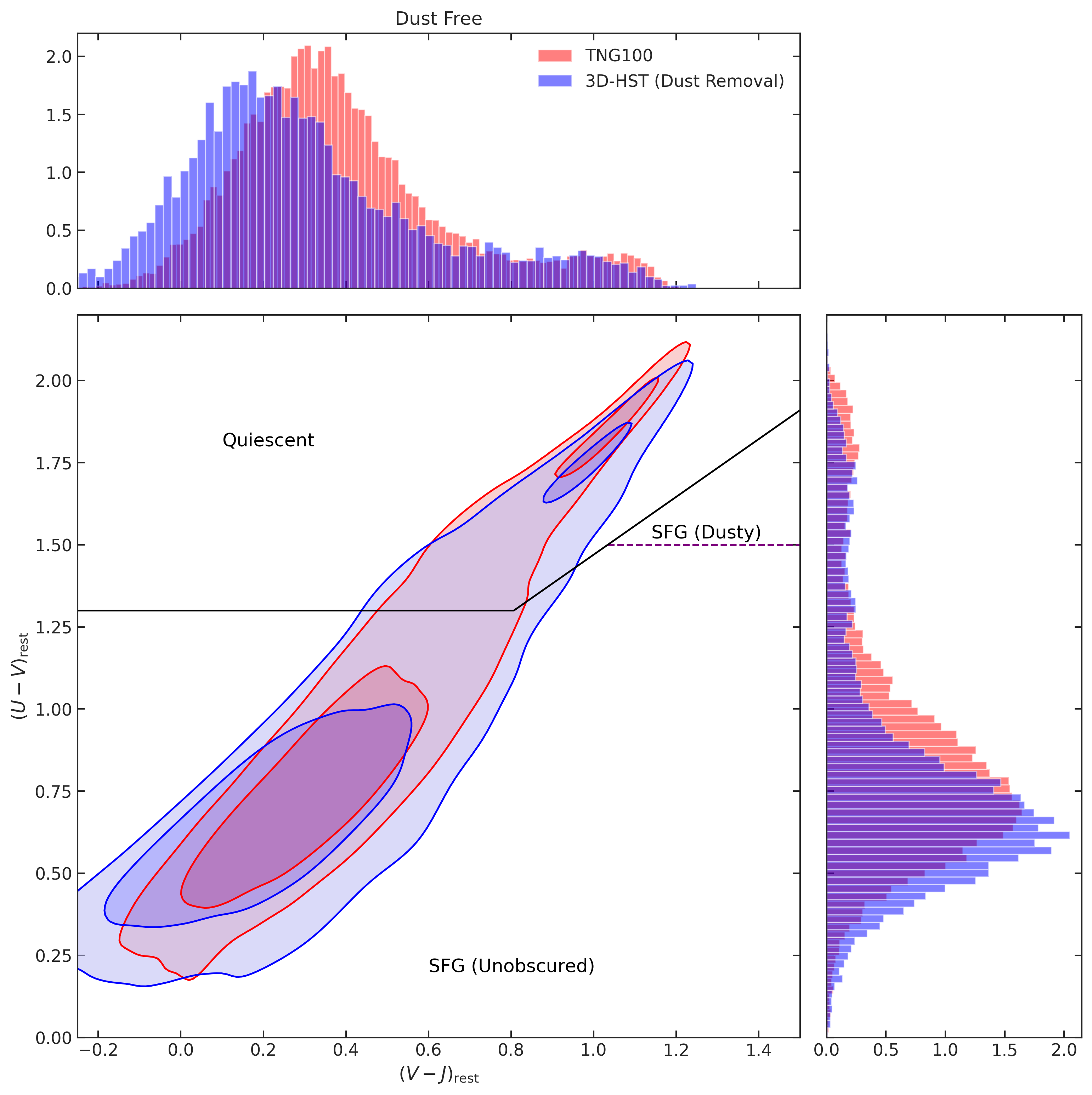}
    \caption{Dust-free UVJ diagram for TNG100 vs 3D-HST. The distribution of dust-free colors from TNG100 at $z=1$---calculated for each galaxy by summing the fluxes from every star particle within $2r_{1/2}$ to mimic the setup in \cite{Donnari2019}--is shown in red.  The distribution of dust-free colors of 3D-HST galaxies (whose calculation is described in \S \ref{subsec:meas3dhst}) is shown in blue. The 68\% (darker) and 95\% (lighter) contours are shaded. See \S \ref{subsec:meastng} for more details on the procedure to calculate colors. The boundaries between quiescent and star forming are taken from \cite{Whitaker2011}, with modifications at the upper right corner inspired by \cite{Belli2019}, while the boundary between dusty and unobscured is from \cite{Fumagalli2014}. The observed galaxies tend to have slightly bluer colors than the simulated galaxies, especially $V-J$.}
    \label{fig:uvj_dust_free}
\end{figure*}

In Figure \ref{fig:uvj_dust_free}, we show the distribution of dust-free TNG100 UVJ colors in red (whose calculation was described in \S \ref{subsec:meastng}). We juxtapose the distribution of colors from the dust-free \prosp 3D-HST galaxies in blue (discussed in \S \ref{subsec:meas3dhst}). 

The boundaries between quiescent and star forming are taken from \cite{Whitaker2011}, with modifications at the upper right corner inspired by \cite{Belli2019}, while the boundary between dusty and unobscured star forming galaxies (DSFGs and USFGs) is from \cite{Fumagalli2014}.

In the top and right panels, we show the 1-D projected distributions of $V-J$ and $U-V$, respectively, for both sources. The \prosp 3D-HST galaxies tend to have slightly bluer colors, especially $V-J$, which we discuss briefly in \S \ref{sec:disc}.

\begin{figure*}
    \centering
    \includegraphics[scale=0.5]{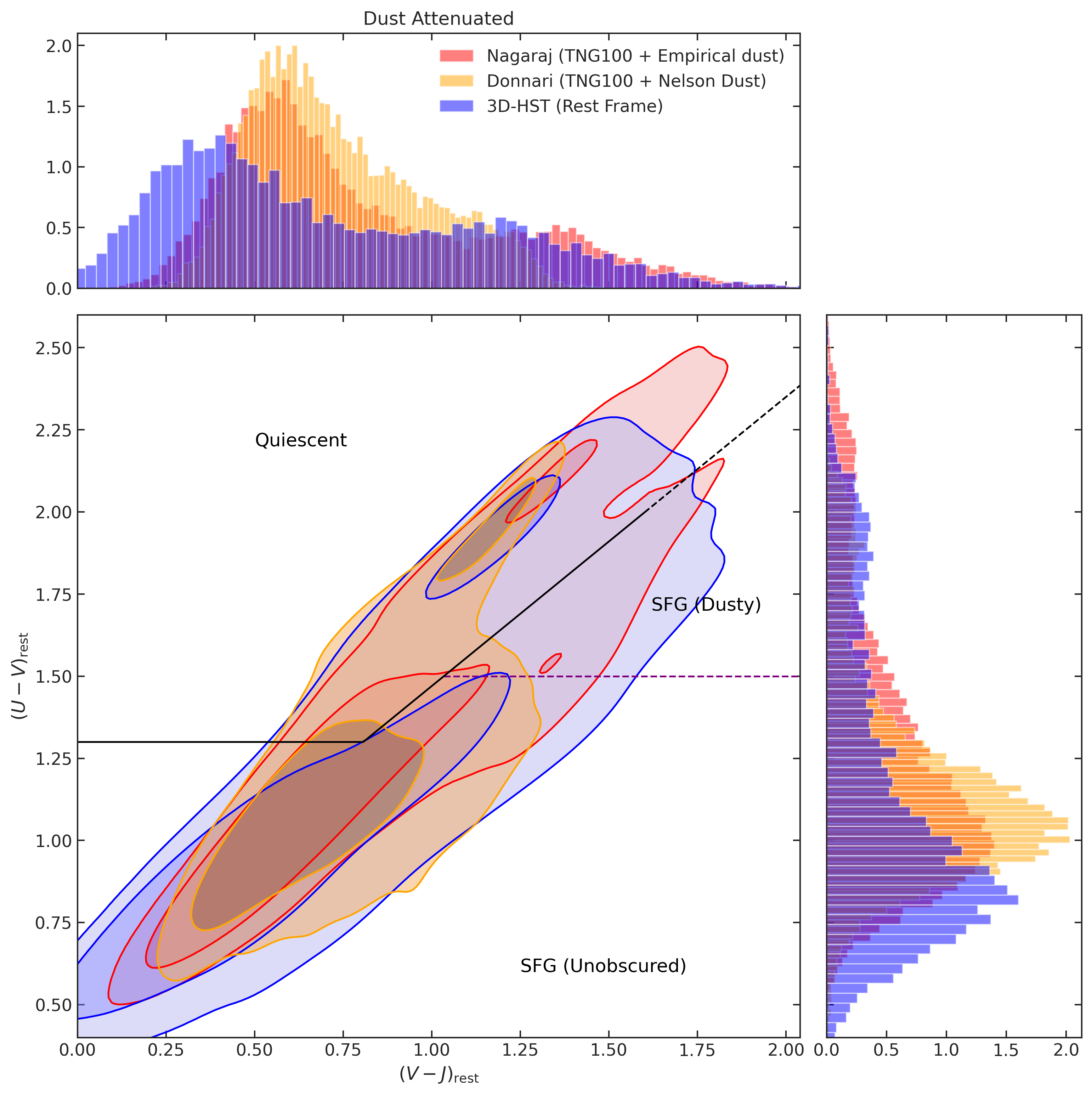}
    \caption{UVJ diagram for TNG100 vs 3D-HST (including dust). The distribution of values from this work are shown in red, 3D-HST galaxies in blue, and those from \cite{Donnari2019} in orange, with 68\% (darker) and 95\% (lighter) contours shaded. Our dust model yields a longer quiescent strip of galaxies as well as many more dusty star forming galaxies (DSFGs) than the resolved \cite{Nelson2018} model used by \cite{Donnari2019}, thereby giving it greater resemblance to the observed UVJ distribution.}
    \label{fig:uvj_dust}
\end{figure*}

Next, we compare the colors with dust included to \prosp rest-frame colors in Figure \ref{fig:uvj_dust}. Once again, the distribution of values from this work are shown in red and that of observed galaxies in orange. In addition, we show the distribution of colors for \cite{Donnari2019} using the theoretical resolved dust model (C) from \cite{Nelson2018} in orange.

The differences between the TNG color distributions measured via two different dust models is quite clear. In particular, we have a much more extended quiescent strip at high $V-J$, a larger range of unobscured star-forming galaxies, as well as a more populated DSFG region. Indeed, the \cite{Donnari2019} distribution includes almost no DSFGs, while we can see they make up a prominent part of the 3D-HST galaxies. 

On the other hand, the \cite{Donnari2019} results have a wider distribution of unobscured star forming galaxies (USFGs) than in this work when $0.5\lesssim V-J \lesssim 1.3$, which is in better agreement with the 3D-HST colors. 

For both TNG efforts, the locus of USFGs is slightly shifted to redder colors than in 3D-HST. However, this likely reflects the same difference in the dust-free colors (Figure \ref{fig:uvj_dust_free})---rather than limitations in the dust models. We discuss this notion further in \S \ref{sec:disc}. Overall, our results are closer to the 3D-HST rest-frame colors than are those from \cite{Donnari2019}.

\section{Discussion} \label{sec:disc}


We now examine the origins of the differences between the distribution of dust-free colors for TNG100 $z=1$ and 3D-HST $0.8\leq z \leq 1.2$ galaxies. We observe in Figure \ref{fig:uvj_dust_free} that the observed galaxies have a slightly bluer locus of points, with a greater effect in $V-J$ than $U-V$. In addition, they occupy a larger part of parameter space. 

\begin{figure*}
    \centering
    \resizebox{\hsize}{!}{
    \includegraphics{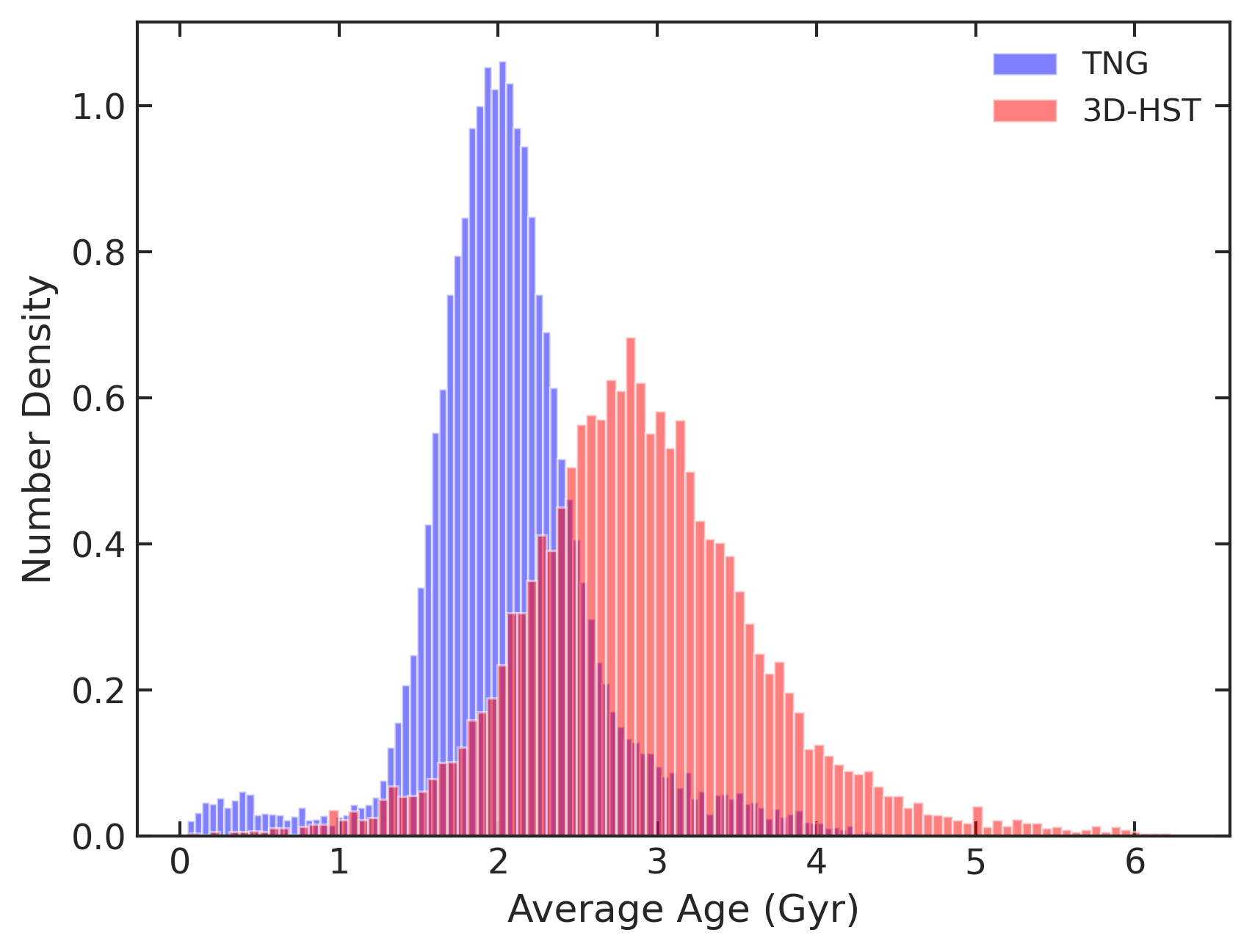}
    \includegraphics{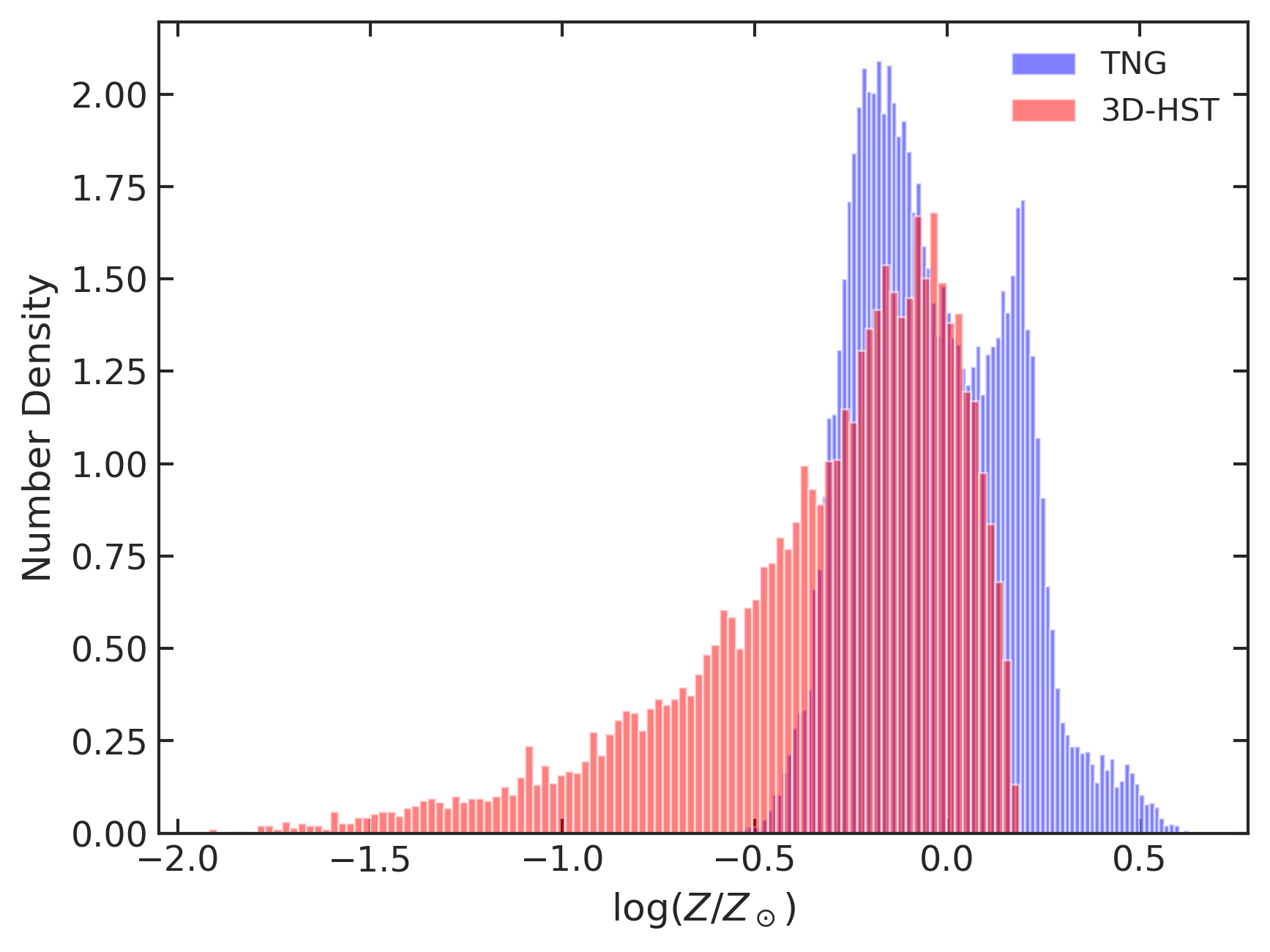}}
    \caption{Distributions of average stellar population age (left) and stellar metallicity (right) for TNG simulated galaxies at $z=1$ (blue) vs 3D-HST galaxies at $0.8\leq z \leq 1.2$. The wider distributions of observed galaxies for age and metallicity may contribute to their larger distribution of UVJ colors. Also, we observe that on average, TNG galaxies have younger stars and higher metallicities. The combination of these two differences can help explain the bluer 3D-HST colors. Higher metallicities signify redder colors, while younger populations lead to bluer colors (opposite effect). The balance of these effects as a function of wavelength can result in the disparity we observe. We offer a more detailed explanation in the text.}
    \label{fig:prosptng}
\end{figure*}

In Figure \ref{fig:prosptng}, we show the distributions of average stellar population age (left) and stellar metallicity (right) for the TNG simulated galaxies at $z=1$ (blue) and 3D-HST galaxies at $0.8 \leq z \leq 1.2$ (red). We immediately notice that the observed galaxies occupy wider distributions in age and metallicity as well as systematically older and more metal-poor populations. 



The larger spread in properties for the observed galaxies is presumably in part due to the wider redshift range $0.8 \leq z \leq 1.2$, a range chosen to include a large enough sample ($9,000$) of \prosp galaxies, compared with a precise $z=0.9973$ for TNG100. However, we find this to be a nearly negligible effect. When we change the range from $0.8 \leq z \leq 1.2$ to $0.9 \leq z \leq 1.1$, the means and standard deviations of the \prosp $U-V$ and $V-J$ distributions are shifted by less than $1\%$. This suggests that the wider \prosp age and metallicity distributions, which are similarly unchanged for the two redshift ranges, are the primary cause of the spread. 

Another important consideration is the fact that the dust-free 3D-HST galaxy colors are not ``directly'' observed but rather calculated with an SED fitting code. Therefore, they are subject to modeling uncertainties and not just galaxy variations. We find that this uncertainty is on the order of $0.1$ magnitudes. By employing our dust attenuation model to re-sample the \prosp posterior distribution (\S \ref{subsec:meas3dhst}), we get a $\sim 15\%$ reduction in the standard deviation of the uncertainties over all galaxies: in other words, we reduce the influence of outliers with very large uncertainties.


The systematically bluer 3D-HST colors, primarily $V-J$, can be explained by the older, more metal-poor stellar populations. Lower metallicities yield bluer colors from UV through NIR while older populations lead to redder colors. Both of these effects are functions of wavelength, meaning the balance between the opposing effects is different for $U-V$ and $V-J$. 



In Figure \ref{fig:uvj_dust}, we compare the dust-included UVJ colors from this work, \cite{Donnari2019}, and 3D-HST rest-frame colors. As shown in \S \ref{sec:res}, the TNG UVJ diagram we calculate in this work is more similar to the 3D-HST observed diagram than is that derived by \cite{Donnari2019}. This is especially true in the higher number densities of dusty star forming galaxies (DSFGs), which are almost completely absent in the \cite{Donnari2019} diagram. The lack of DSFGs has also been noted in the \texttt{MUFASA} simulation \citep{Dave2017} and is potentially related to resolution limitations.

Nevertheless, important differences remain between our TNG UVJ diagram and the 3D-HST rest-frame colors. For example, the distribution we derive is more compact, with a smaller portion of color-color space being occupied by unobscured star forming galaxies. In addition, it is slightly redder than the observed distribution. 

One of the reasons for this phenomenon is the intrinsic dust-free color differences. The 3D-HST distribution is considerably wider than TNG100. Naturally, that intrinsic thickness will be imprinted on the dust-included colors. Similarly, the redder dust-free TNG colors help lead to redder dust-included colors.



In addition to the dust-free considerations, we can attribute some of the differences to contrasts between the population dust model and the underlying data on which it was built. When we replace the individually fit dust parameters (used to reproduce the rest-frame spectra) with our dust attenuation model (not portrayed in the paper), the resulting distribution is more compact and very slightly redder than the observed distribution, suggesting that there are still some limitations in the dust model that affect the UVJ distribution. 

Perhaps the most important point is that in the 5-D linear interpolation model, we are able to include only five grid points per dimension ($3,125$ model parameters total) to achieve convergence. Therefore, small-scale variations in the dust attenuation function will be smoothed over. In addition, we assume a single intrinsic scatter in the relationship for simplicity; in reality, the scatter is likely a function of parameters. Finally, there may be additional variables needed to fully explain the scatter in attenuation curves. Nevertheless, we have demonstrated in both Paper~I and this work the prowess of the population dust attenuation model in its high level of statistical accuracy and ability to provide a more reliable interface between simulations and observations.

\begin{figure*}
    \centering
    \includegraphics{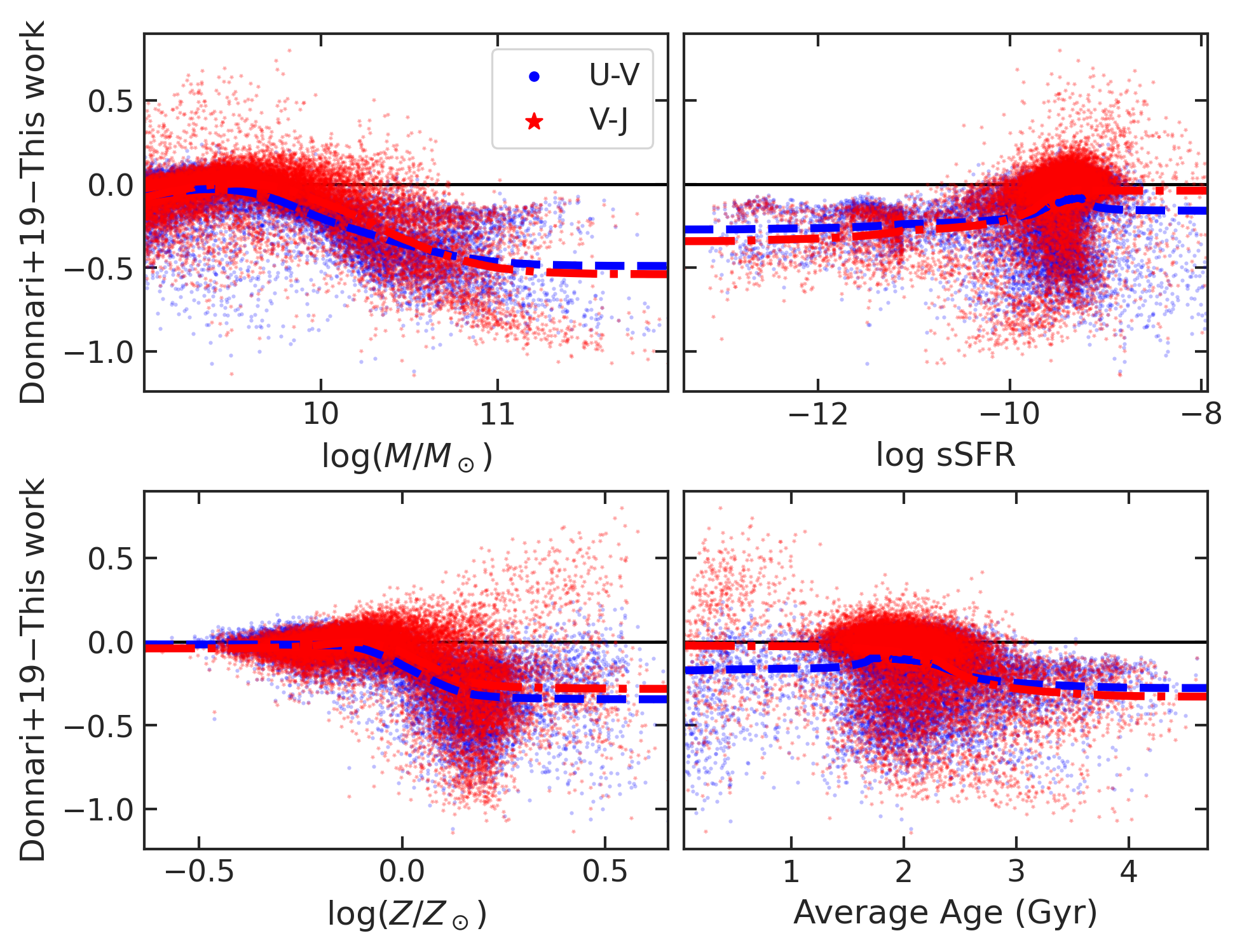}
    \caption{Galaxy-by-galaxy differences between dust-attenuated UVJ colors for TNG100 galaxies derived in this work and \cite{Donnari2019} as a function of stellar mass (top left), sSFR (top right), metallicity (bottom left), and average age of stellar populations (bottom right). Very minor differences between the dust-free UVJ colors are removed to create a fair baseline for the comparisons. We see that the UVJ differences are systematically negative for high stellar mass, low sSFR, high metallicity, and/or older galaxies. In other words, \cite{Donnari2019} predicts bluer galaxies in that part of parameter space. If the effective attenuation curve obtained by \cite{Nelson2018} based on the simulation particle data were replaced by an effective dust screen, the predicted attenuation curve would either have a lower optical depth or be flatter than our empirical curve for such galaxies.}
    \label{fig:uvjdiff}
\end{figure*}

Analyzing how the differences between the UVJ colors derived in this work and \cite{Donnari2019} vary as a function of physical properties like stellar mass will help us understand the underlying differences in the dust models used. In Figure \ref{fig:uvjdiff}, we show the difference in $U-V$ (blue) and $V-J$ (red) as a function of stellar mass (top left), sSFR (top right), metallicity (bottom left), and average age of stellar populations (bottom right). To remove traces of differences in forward-modeling the TNG SEDs, we subtract the differences in derived dust-free colors from the result. For example, for $U-V$, the quantity plotted is $(U-V)_{\rm Don} - (U-V)_{\rm Nag} - \left[ (U-V)_{\rm{Don,~dust-free}} - (U-V)_{\rm{Nag,~dust-free}} \right]$.

From Figure \ref{fig:uvjdiff}, we observe that for both $U-V$ and $V-J$, the color residuals are systematically negative for high stellar mass, low sSFR, high metallicity, and/or older galaxies. This equates to a prediction of bluer galaxies in that part of parameter space from the resolved dust model of \cite{Nelson2018}. Comparing a resolved and unresolved dust model is difficult as the effects of geometry are complex, but we can make a simple statement: if the resolved model were replaced by an effective dust screen, the resulting attenuation curve would either have a smaller optical depth and/or be flatter than our empirical attenuation model for high-mass, high-metallicity, low-sSFR, and/or old galaxies.

\newpage
\section{Conclusion} \label{sec:conc}

We have created a population Bayesian model of dust attenuation (Paper~I). In this paper, we detail its usage in theoretical scenarios and demonstrate its success by applying it to TNG100 galaxies at $z=1$ and comparing the resulting UVJ colors to both previous efforts to model the effects of dust in these galaxies \citep{Nelson2018,Donnari2019} and rest-frame colors of 3D-HST galaxies. 

Overall, our population Bayesian dust attenuation model results in a more observation-like UVJ diagram for TNG100 galaxies with stellar masses $M\geq 10^9 \msun$ than previous efforts. 

Furthermore, by using the population model to reweight posterior samples of individual galaxy SED fits (hierarchical shrinkage), we are able to calculate more reliable dust-free rest-frame colors for observed galaxies and thus uncover more fundamental differences between TNG galaxies and the observed universe. In particular, we find that on average, TNG galaxies have a 29\% younger and 28\% narrower (stellar population) age distribution and a 45\% narrower and $0.28$ dex more metal-rich stellar metallicity distribution, leading to a slightly redder dust-free distribution for TNG100 galaxies than \prosp 3D-HST galaxies.

Dust plays a major role in determining the spectra of galaxies, but, as of yet, its effects cannot be reproduced from first principles in galaxy formation and evolution simulations. Compounding the issue, its effects on integrated galaxy spectra are hard to distinguish from those of metallicity and age, and thus confound comparisons with observations. By using a statistically rigorous empirical approach to create our model, we are able to provide a simple but accurate interface between theoretical (dust-free) models and observed SEDs. Through the efforts described in this paper, we are able to better understand the limitations of such models (e.g., issues in the treatment of chemical evolution) without most of the complications from dust.


\acknowledgments

We would like to thank Sandro Tacchella for computing star formation histories, metallicity histories, and stellar masses of TNG100 subhalos for us. We would also like to thank Dylan Nelson for providing us with UVJ colors used in \cite{Donnari2019} with and without dust for TNG100 subhalos. We also thank Julianne Dalcanton for insightful conversations.

This material is based upon work supported by the National Science Foundation Graduate Research Fellowship under Grant No. DGE1255832. Any opinion, findings, and conclusions or recommendations expressed in this material are those of the authors(s) and do not necessarily reflect the views of the National Science Foundation. The Flatiron Institute is supported by the Simons Foundation.

%

\vspace{5mm}
\facilities{HST (WFC3), Spitzer (MIPS)}


\software{\texttt{pymc3} \citep{Pymc3}, AstroPy \citep{Astropy2018}, SciPy \citep{Scipy2020}, FSPS \citep{Conroy2009,Conroy2010SPSM}, \prosp \citep{Leja2017,Johnson2021} }




\vspace{4cm}

\bibliography{sample63}{}
\bibliographystyle{aasjournal_mod}



\end{document}